\documentclass[prb,twocolumn]{revtex4-1} 

\usepackage{amsmath}  
\usepackage{amsfonts} 
\usepackage{graphicx} 

\begin{document}

\title{The quantum character of buckling instabilities in thin rods}

\author{T. A. Engstrom}
\email{tyler.engstrom@gmail.com} 
\affiliation{Department of Physics, Hobart and William Smith Colleges, Geneva, NY 14456, USA}
\affiliation{Department of Physics and Astronomy, Colgate University, Hamilton, NY 13346, USA}

\date{\today}

\begin{abstract}
Here the buckling of inextensible rods due to axial body forces is mapped to 1d, nonrelativistic, time-independent quantum mechanics. Focusing on the pedagogical case of rods confined to 2d, three simple and physically realizable applications of the mapping are given in detail; the quantum counterparts of these are particle in a box, particle in a delta-function well, and particle in a triangular well. A fourth application examines the buckling counterpart of a quantum many-body problem (in the Hartree approximation). Through a fifth application, given in the form of an exercise, the reader can explore the surprising consequences of adding a second transverse dimension to the rod buckling problem and imposing periodic boundary conditions. 
\end{abstract}

\maketitle

\section{Introduction} 

Energy quantization, normalization, and observable consequences of relative phase differences are usually regarded as concepts belonging to the realm of quantum mechanics. Yet these same concepts also apply to elastic buckling instabilities in thin rods, the theory of which dates back to the days of Euler,\cite{gautschi08} but is not part of standard physics curriculum. Very accessible introductions to buckling and its applications can be found in the physics education literature,\cite{taberlet17,casey93,denny19, mills60} while a classic treatise on the subject is given by Landau \& Lifshitz.\cite{landau86} Several prior analogies have been made between elastic buckling problems and quantum mechanical problems,\cite{hui-chuan85, hui-chuan87, odijk86, odijk98, hansen99, ohayre16} but a general and unifying framework for such analogies is absent. In some of the abovementioned works, the wavefunction-like entity is said to be the shape (deflection) of the buckled rod,\cite{hui-chuan85, hui-chuan87, ohayre16} in others the slope (tangent vector) of the rod,\cite{odijk86, odijk98} and yet in another the curvature of the rod.\cite{hansen99} All place restrictions on the form of the potential energy-like entity, it being either a constant,\cite{hui-chuan85, hui-chuan87, ohayre16} or of a harmonic oscillator form,\cite{odijk86, odijk98} or a symmetric function with respect to the midpoint of the rod.\cite{hansen99} Here we present a formal mapping between rod buckling in 2d and time-independent quantum mechanics in 1d that is considerably more general than those analogies suggested before. We find that if the wavefunction-like entity is taken to be the slope of the rod, then the normalization condition maps directly to an inextensibility constraint for the rod, and the potential energy function maps to an arbitrary body force acting parallel to the rod.

Let us start by briefly reviewing deformations of rods.\cite{landau86} An elastic rod can deform by bending, by stretching or compressing lengthwise, and by torsion, while still remaining a rod. We will specialize to rods with small, smooth deformations confined to a plane (in other words, nearly straight rods in 2d), and ignore torsion. If such a rod is oriented along $\hat{x}$, with transverse deflection $u(x)$ much smaller than the rod's length, and axial strain $\epsilon(x)\ll1$, the deformation energy of the rod is given by
\begin{equation}
H = \frac{\kappa}{2} \int dx \bigg(\frac{d^2u}{dx^2}\bigg)^2  +  \frac{\mu}{2} \int dx \; \epsilon^2. \label{rod_energy}
\end{equation}
Here $\kappa$ is called the bending modulus and $\mu$ is called the stretching modulus. For a rod with an equiaxed cross section, these scale as $\kappa\sim A^2Y$ and $\mu\sim AY$, where $Y$ is Young's modulus and $A$ is the cross sectional area. Thus if the rod is very thin ($A\to0$), its resistance to stretching and compressing is much greater than its resistance to bending. Biopolymers, carbon nanotubes, and certain other filamentous molecules are examples of very thin rods, and they are often modeled by removing the second term in Eq.~(\ref{rod_energy}) and replacing it with an inextensibility constraint. This is the main idea of the ``worm-like chain" (WLC) model,\cite{kratky49, broedersz14, marantan18, devenica16} which we will again encounter in the following analysis.

\section{Formal mapping}

Suppose the rod is subjected to a net contact and/or body force $T(x)$ that acts parallel to $\hat{x}$. To leading order, the axial stress in the rod is $T(x)/A$ and the axial strain is $T(x)/\mu$. If the rod is in equilibrium, any small section of it must obey the equation of local moment balance $dM=Tdu$, where $M(x)=\kappa u''$ is the bending moment. Dividing both sides by the length of the section, $dx$, one obtains the third-order equation of shearing force equilibrium,\cite{landau86}
\begin{equation}
\kappa\frac{d^3u}{dx^3} = T(x)\frac{du}{dx}. \label{shearing_force_equil}
\end{equation}
Regions of $T>0$ correspond to tension, while regions of $T<0$ correspond to compression. A typical application of Eq.~(\ref{shearing_force_equil}) is to ``self-buckling," which refers to a vertical column of height $h$ that buckles under its own weight: $T(x)=-\sigma(h-x)$, where $\sigma$ is the weight per unit length.\cite{landau86, taberlet17} By a change of variable $w\equiv du/dx$, Eq.~(\ref{shearing_force_equil}) takes the form of the 1d time-independent Schr\"odinger equation,
\begin{equation}
\kappa \frac{d^2w}{dx^2} = T(x)w. \label{shearing_force_equil_2}
\end{equation}
This is similar in appearance to the second-order equation of moment equilibrium obtained by integrating Eq.~(\ref{shearing_force_equil}) for the special case of constant $T$. That result, $\kappa u''=Tu$, is commonly known as the Euler buckling equation. It should be clear, however, that  Eq.~(\ref{shearing_force_equil_2}) is more general than the Euler buckling equation, and it also has a different physical meaning. Boundary conditions for Eqs.~(\ref{shearing_force_equil}) and~(\ref{shearing_force_equil_2}) typically involve hinged or clamped rod ends, and solutions $u(x)$ and $w(x)$ to such boundary value problems describe unstable equilibrium configurations of the rod, i.e., buckled configurations.

Letting $\mathbf{r}$ be the 2d displacement vector that locates one end of the rod with respect to the other, we define a ``projected length" of the rod as $L\equiv|\hat{x}\cdot\mathbf{r}|$. In a buckled configuration, the rod's contour length $L_C$ exceeds $L$ by an amount
\begin{equation}
L_C-L=\int_0^Ldx\big(\sqrt{1+w^2}-1\big).
\end{equation}
Small, smooth deformations imply $w(x)\ll1$ everywhere, permitting Taylor expansion of the square root. Doing this and rearranging terms, we have
\begin{equation}
L_C=L + \frac{1}{2}\int_0^L dx\,w^2.\label{DeltaL}
\end{equation}

Now an \emph{inextensible} rod is one that cannot change its contour length. Geometry dictates that any change made to the projected length must be absorbed entirely into the buckling amplitude, via the second term on the right hand side of Eq.~(\ref{DeltaL}). To see this, suppose the rod is initially straight with $L=L_C$, and the projected length is subsequently reduced to $L=L_C-\Delta L$; we must then have $\frac{1}{2}\int_0^L dx w^2 = \Delta L$. Defining a relative change in projected length $\gamma \equiv \Delta L/L\ll1$ (not to be confused with the axial strain in the rod, which is zero), we can write the inextensibility constraint as
\begin{equation}
\int_0^L dx \bigg(\frac{w}{\sqrt{2\gamma L}}\bigg)^2 = 1. \label{delta_L_2}
\end{equation}
So for an inextensible rod, i.e. a WLC, not only does the slope $w(x)$ satisfy a Schr\"odinger-like equation, it satisfies a geometrical constraint that is reminiscent of normalization.

Introducing a rescaled slope $W(x)\equiv  w(x)e^{i\phi}/\sqrt{2\gamma L}$ that is dimensionally consistent with a 1d quantum mechanical wavefunction, where $\phi$ is an arbitrary constant phase angle, Eqs.~(\ref{shearing_force_equil_2}) and~(\ref{delta_L_2}) become
\begin{eqnarray}
\frac{d^2W}{dx^2} - \frac{T(x)}{\kappa} W &=& 0,\label{W1}\\
\int dx\,|W|^2 &=& 1.\label{W2}
\end{eqnarray}
The integration is over the projected length of the WLC. Evidently the problem of generalized buckling instabilities in 2d WLCs maps to 1d, nonrelativistic, time-independent quantum mechanics according to
\begin{eqnarray}
\frac{w(x)}{\sqrt{2\gamma L}} \hskip 1em &\mapsto& \hskip 1em \psi(x), \label{Wtopsi}\\
-\frac{T(x)}{\kappa} \hskip 1em &\mapsto& \hskip 1em \frac{p^2(x)}{\hbar^2}. \label{TtoE}
\end{eqnarray}
Here $\psi(x)$ is a real, normalized eigenstate of the time-independent Schr\"odinger equation belonging to eigenenergy $E$, and $p(x)=\sqrt{2m[E-V(x)]}$ is the semi-classical momentum. Regions of compression of the WLC map to classical regions: $E-V(x)>0$, while regions of tension map to nonclassical regions: $E-V(x)<0$. A neutral ``surface" of the WLC maps to a classical turning point in the quantum problem. A boundary condition in which the WLC is clamped parallel to $\hat{x}$ (but the clamp can slide transversely) maps to a boundary condition in which $\psi$ vanishes.

Further aspects of the mapping are obtained from energy considerations. Let $H_{bend}$ denote the first term on the right hand side of Eq.~(\ref{rod_energy}). Substituting $w'=u''$ and inserting Eq.~(\ref{Wtopsi}), we find
\begin{equation}
\frac{H_{bend}}{\kappa\gamma L} \hskip 1em \mapsto \hskip 1em \frac{\langle \hat{p}^2\rangle}{\hbar^2}, \label{H_bend}
\end{equation}
where $\langle\hat{p}^2\rangle=\hbar^2\int dx \,|d\psi/dx|^2$ is the expectation value of the squared momentum operator in the state $\psi$ (after integrating by parts). Notice that the choice $\kappa\gamma L=\hbar^2/2m$ maps $H_{bend}$ directly to the expectation value of kinetic energy. Next we observe that the work done on the rod by the body force is $U=-(1/2)\int dx \,Tw^2$. Inserting Eqs.~(\ref{Wtopsi}) and~(\ref{TtoE}) reveals
\begin{equation}
\frac{U}{\kappa\gamma L} \hskip 1em \mapsto \hskip 1em \frac{\langle {p}^2\rangle}{\hbar^2}.\label{U}
\end{equation}
This time, the expectation value is of the squared \emph{semi-classical} momentum: $\langle p^2\rangle=\int dx\, p^2|\psi|^2$. Thus, the statement of energy conservation in the buckling problem, $H_{bend}=U$, is akin to multiplying the time-independent Schr\"odinger equation on the left by $\psi^*$ and integrating.  

In the following section we examine four sample applications of the mapping that span a wide range of qualitative behaviors. The first two involve only contact forces, while the latter two involve body forces.

\section{Applications of the mapping}

\subsection{Particle in a box}

\begin{figure}[t]
\centering
\includegraphics[width=0.485\textwidth]{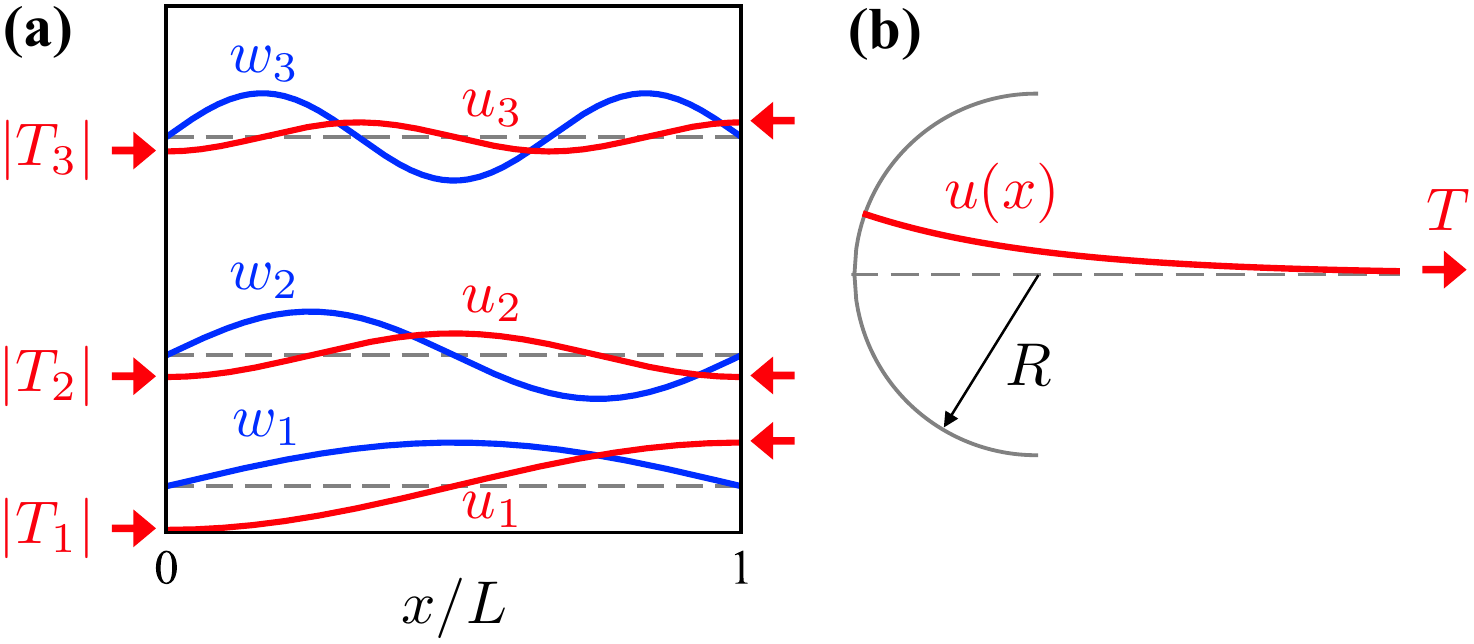}
\caption{Modes of buckling instability analogous to particle in a box eigenstates and the delta-function well bound state, with all boundary conditions involving a sliding clamp. Panel (a): the WLC shapes $u_n(x)$ are shown in red, and the derivatives of these shapes are the ``eigenslopes" $w_n(x)$, shown in blue. Notice that while the $w_n$ have constant amplitude (assuming a fixed value of $\gamma$), the $u_n$ amplitude scales as $1/n$, as required by the inextensibility constraint. The compressive force required to generate the $n^{th}$ buckling mode is the ``eigenload" $|T_n|$, and the dashed lines indicate the spacing between eigenloads. Panel (b): the single mode of buckling instability for a tensioned WLC whose left end clamp is constrained to slide along a circular track.\cite{misseroni15} In both panels the WLC shapes and slopes are greatly exaggerated for clarity. \label{euler_delta}}
\end{figure}

The buckling problem analogous to a particle in a 1d infinite square well of width $L$ is
\begin{equation}
\kappa w''=-|T|w, \hskip 2em  w(0)=w(L)=0, \label{PinB_analog}
\end{equation}
where $T=$ constant. Physically, this represents a WLC compressed from its endpoints; the ends are clamped but the clamps are free to slide transversely. Since the eigenvalues of Eq.~(\ref{PinB_analog}) are compressive loads and the eigenfunctions describe the WLC's slope, we use the more descriptive names ``eigenloads" and ``eigenslopes." These are given by $|T_n|=n^2\pi^2\kappa/L^2$ and $w_n(x)=2\sqrt{\gamma}\sin(n\pi x/L)$, respectively, and the first few are shown in Fig.~\ref{euler_delta}(a). 
%
%
One can easily verify that the eigenslopes satisfy the inextensibility constraint (Eq.~(\ref{delta_L_2})) and the mapping to the normalized $\psi_n$ (via Eq.~(\ref{Wtopsi})). The eigenvalues of the two problems are related by $2mE_n/\hbar^2=|T_n|/\kappa$, consistent with Eq.~(\ref{TtoE}). 

What are the consequences of choosing $\kappa\gamma L=\hbar^2/2m$, which maps the bending energy directly to the kinetic energy, as mentioned above? One consequence would be that the buckling force is $|T_n| = E_n/\gamma L$, much larger than the force required to adiabatically change the width of the well $-dE_n/dL=2E_n/L$. Is that problematic? No, it appears to be reasonable behavior given that adiabatically changing the width of the well corresponds to changing the projected length of the already-buckled WLC. (Engineers consider buckling to be a mode of failure because a generic rod can support a much greater axial load prior to buckling than after it has buckled, and a WLC is no exception to this rule.) Another consequence is thermodynamic in nature. A fundamental property of a WLC is its persistence length $l_p=2\kappa/\tau$ (in 2d), defined as the decay length of the tangent-tangent correlation function $\langle \hat{t}(s)\cdot\hat{t}(s') \rangle = \exp(-|s-s'|/l_p)$. Here $\tau$ is the Boltzmann constant times temperature, and $\hat{t}(s)$ is the unit vector tangent to the WLC at distance $s$ measured along its contour length. Since the mapping is valid only for small transverse deflections of the WLC, we must be confined to the ``stiff" regime $L\lesssim l_p$, i.e., the low temperature regime of the WLC.\cite{broedersz14} Under this restriction, we would have
$\sqrt{2\pi\gamma l_pL} = \lambda_{th}$, where $\lambda_{th}$ is the thermal average wavelength of the particle in a box, i.e., a 1d ideal gas of density $L^{-1}$. Inserting the condition $L\lesssim l_p$ into the last equation indicates the ideal gas would be in the density regime $L^{-1} \gtrsim \sqrt{2\pi\gamma} \,\lambda_{th}^{-1}$, where $\lambda_{th}^{-1}$ is known as the quantum concentration.\cite{kittel80} However, since $\gamma\to0$, this is not actually a restrictive condition; the cold WLC picture would hold regardless of whether the ideal gas is in the quantum or classical regime (density above or below $\lambda_{th}^{-1}$, respectively).

\subsection{Particle in a delta-function well}

The time-independent Schr\"odinger equation
\begin{equation}
\frac{-\hbar^2}{2m}\frac{d^2\psi}{dx^2} - g\delta(x)\psi =E\psi, \label{delta_fct}
\end{equation}
can be recast as separate boundary value problems for each half-space. For the positive half-space,
\begin{equation}
\psi''-k^2\psi=0, \hskip 1em \psi(\infty)=0, \hskip 1em  \psi'(0)=\frac{-1}{R}\psi(0), \label{delta_bvp}
\end{equation}
where $k^2=2m|E|/\hbar^2$, $R=\hbar^2/mg$, and the second boundary condition comes from integrating Eq.~(\ref{delta_fct}) across an infinitesimal region centered on the origin. The sole bound state solution is $\psi(x)=\sqrt{1/R}\,e^{-x/R}$, and the energy of this state is $E=-mg^2/2\hbar^2$.

The analogous buckling problem has been described by Misseroni, et al.\cite{misseroni15} Clamp one end of a rod and constrain the clamp to slide along a circular path having radius $R$, then pull on the other end (see Fig.~\ref{euler_delta}(b)). Here the rod is an infinitely long WLC, so the boundary value problem is
\begin{equation}
w''-q^2w=0, \hskip 1em w(\infty)=0, \hskip 1em  w'(0)=\frac{-1}{R}w(0), \label{euler_tension}
\end{equation}
where $q^2=T/\kappa>0$. The WLC will remain straight until the tension reaches a critical value $T=\kappa/R^2$, at which point it will deflect and acquire slope $w(x)=\sqrt{2\gamma L/R}\,e^{-x/R}$. (Note $\gamma L/R\to0$ is implicit here.) Just as there is only one bound state for the delta-function well, there is only one ``buckling" mode for the tensioned rod. The binding energy and buckling force are related by $E=-\hbar^2T/2m\kappa$, again consistent with Eq.~(\ref{TtoE}).

\subsection{Particle in a triangular well}

\begin{figure}[t]
\centering
\includegraphics[width=0.485\textwidth]{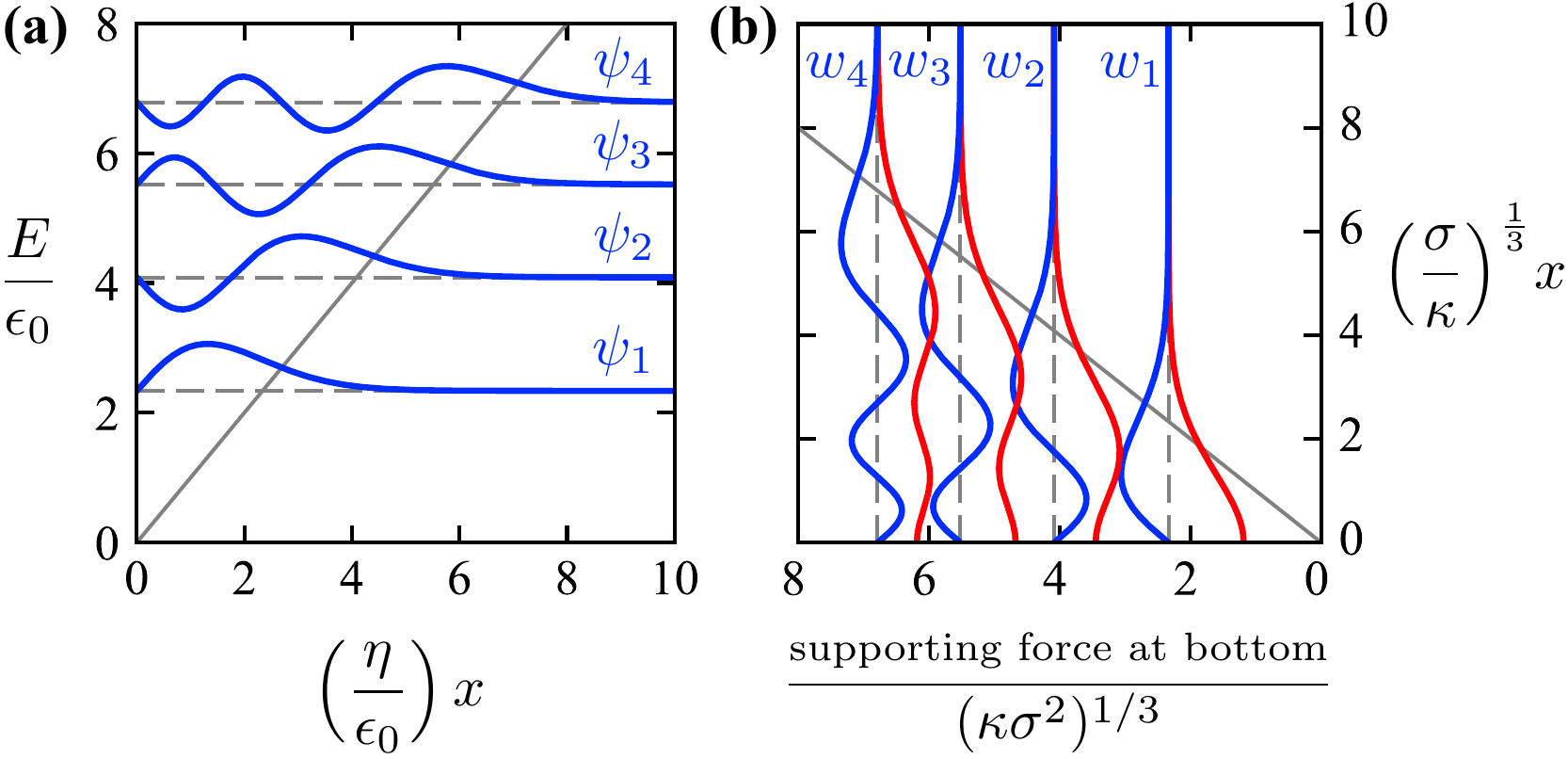}
\caption{Particle in a triangular well maps to a variant of self-buckling in which there is a supporting force at the bottom of a vertical WLC and a suspending force at the top, and the sum of these two forces equals the weight of the WLC. Panel (a): the first four normalized eigenstates, vertically shifted to their respective energy levels, for a fixed value of $\eta/\epsilon_0$. The potential $V(x)$ is the diagonal grey line, plotted in units of $\epsilon_0$. Panel (b): identical to left panel, but rotated 90 degrees. Here the blue lines are the WLC slopes $w_n(x)$, where $n$ indexes the mode of buckling instability. The corresponding WLC shapes are shown in red, and found by numerical integration as $u_n=\int dx\,w_n$. As before, the slopes and shapes are exaggerated for clarity, and the boundary conditions involve sliding clamps. Above the diagonal grey line, the WLC is in tension, and below the line it is in compression.\label{triangular_well}}
\end{figure}

While the previous two applications involved only contact forces (applied to the ends of the WLC and transmitted throughout its length as required by force balance), this application involves both contact forces and a body force. First we consider the quantum problem of a particle in a potential well $V(x)=\eta x$ for $x>0$ and $V(x)=\infty$ for $x<0$, where $\eta$ is a constant force. Physically, this could describe an electron near a doped heterojunction,\cite{davies97} or a quantum bouncing ball.\cite{banacloche99} Schr\"odinger's equation is given by
\begin{equation}
\psi'' - \frac{2m\eta}{\hbar^2}\bigg(x-\frac{E}{\eta}\bigg)\psi=0, \hskip 2em \psi(0)=\psi(\infty)=0. \label{triangular}
\end{equation}
The eigenstates (plotted in Fig.~\ref{triangular_well}(a)) are
\begin{equation}
\psi_n(x) = \frac{\sqrt{\eta/\epsilon_0}}{|\textrm{Ai}'(a_n)|}\textrm{Ai}\bigg(\frac{\eta x-E_n}{\epsilon_0}\bigg), \label{airy}
\end{equation}
where Ai$(z)$ and Ai$'(z)$ denote the Airy function and its derivative, $E_n=|a_n|\epsilon_0$,
\begin{equation}
\epsilon_0=\bigg[\frac{(\eta\hbar)^2}{2m}\bigg]^{1/3},
\end{equation}
and $a_n<0$ is the $n^{th}$ zero of the Airy function. The normalization of Eq.~(\ref{airy}) can be verified using an integral identity given by Stern.\cite{stern72}
%
%
The analogous elastic instability problem is a variation on the ``self-buckling" scenario described earlier. Suppose a massive WLC is oriented parallel to a uniform gravitational field. It is both supported from its bottom (at $x=0$) and suspended from its top (at $x=L$), such that it has a neutral ``surface" at some height $x_0$ between 0 and $L$. The axial force is $T(x)=-\sigma(x_0-x)$. Both ends of the WLC are clamped, but the clamps are free to slide transversely as in the previous two applications. For $L\to\infty$, the boundary value problem describing shearing force equilibrium is
\begin{equation}
w''-\frac{\sigma}{\kappa}(x-x_0)w=0, \hskip 2em w(0)=w(\infty)=0.
\end{equation}
This is identical to Eq.~(\ref{triangular}) when $\sigma/\kappa=2m\eta/\hbar^2$, and $x_0=E/\eta$. The classical turning points $E_n/\eta$ in the quantum problem become the neutral surfaces $(x_0)_n$ in the elastic problem (see Fig.~\ref{triangular_well}(b)). The bottom supporting force for the $n^{th}$ buckling mode is given by $T_n(0)=-\sigma(x_0)_n=-|a_n|(\kappa\sigma^2)^{1/3}$, and this maps to the energy of the quantum particle via $p^2_n(0)=2mE_n$, according to Eq.~(\ref{TtoE}).

\subsection{Many interacting particles}
 
What is the buckling instability counterpart of a quantum many-body problem? As a very basic starting point, we show how a Hartree-like term\cite{ashcroft76, blinder65} could arise for a bundle\cite{broedersz14} of interacting WLCs. First we revisit Eq.~(\ref{DeltaL}) and notice that, physically, $w^2(x)/2$ is the ``excess length density," i.e., the fraction of the WLC's total excess length $L_C-L$ found between $x$ and $x+dx$. Suppose the WLC had a charge uniformly spread over its contour length; the charge per unit projected length would be $\sim C+w^2(x)$, where $C$ is a constant. Now consider a bundle of charged WLCs that are all in buckled configurations (but not necessarily the same configuration). If the charged WLCs have 2d electrostatic interactions with one another, then the magnitude of the body force on the $i^{th}$ WLC from all the others is
\begin{equation}
F_i(x) \sim \sum_{j\neq i} \int dx' \frac{\big[C+w_i^2(x)\big]\big[C+w_j^2(x')\big]}{\sqrt{(x-x')^2 + (u_i(x)-u_j(x'))^2}}. \label{force_on_i}
\end{equation}
(Here the subscripts label WLCs, not modes of instability.) Since the $u$'s and $w$'s are small quantities, the transverse component of the body force is small, and Eq.~(\ref{force_on_i}) is well approximated by 
\begin{equation}
T_i(x) \sim \sum_{j\neq i} \int dx' \frac{C\big[C+w_j^2(x')\big]}{|x-x'|}.
\end{equation}
The divergent part of this integral can presumably be discarded, and what remains is a Hartree-like contribution to the total force $T(x)$ exerted on the $i^{th}$ WLC.

\section{Discussion}

We have shown by a formal mapping, and by several applications of the mapping, that key features of nonrelativistic, time-independent quantum mechanics are also contained within a certain class of rod buckling problems. In these problems, the product of a WLC's bending modulus $\kappa$ and change in projected length $\gamma L$ plays the role of $\hbar^2/m$, and the spatial derivative of the WLC's shape plays the role of the normalized wavefunction. The statement of shearing force balance for the WLC is analogous to the statement of energy conservation that is embodied by the Schr\"odinger equation. As mentioned in the Introduction, other quantum-buckling analogies can and have been made using second and fourth-order elastic equations. However, these other analogies do not appear to have the combination of generality (in the sense of accommodating arbitrary $V(x)$) and depth (in the sense that the dependent variable simultaneously satisfies a normalization-like constraint) that is inherent to the third-order equation of shearing force balance.

Generalization of the mapping to higher dimensions is possible in at least three different senses: (1) The inextensible rod can become an inextensible ribbon of arbitrary width (measured perpendicular to the page in Figs.~\ref{euler_delta} and~\ref{triangular_well}). This doesn't change anything in the analysis we've already done, and in fact, by assuming 2d electrostatics in the last application, we've already made use of a ribbon concept. (2) In the context of the first application, the inextensible rod subject to uniaxial compression can become an inextensible sheet subject to biaxial compression in the $x$-$y$ plane, giving rise to independent sinusoidal profiles in the $x$-$z$ and $y$-$z$ planes. (3) The rod might live in 3d space so it has not one but two transverse dimensions into which it can buckle. This latter situation is particularly compelling because the rescaled slope $W(x)$ becomes a 2d vector $\mathbf{W}(x)$, the components of which play the role of the real and imaginary parts of the 1d wavefunction. So the pedagogical mapping given above is actually a special case of a more general isomorphism that exists between WLC buckling and 1d time-independent quantum mechanics. Further details of this isomorphism are given in the Appendix, along with a suggested exercise.

What about time-dependence in the buckling problem --- does it resemble time-dependence in quantum mechanics? The general equation of motion of a vibrating WLC is $\rho\partial_t^2u = -\kappa\partial_x^4u + T\partial_x^2u - \beta u$, where $\rho$ is the WLC's mass per unit length and $\beta$ is the stiffness of a substrate that we have not heretofore considered. In the special case of constant coefficients, taking a spatial derivative of this equation allows us to replace $u$ with $W$, and going to two transverse dimensions further changes $W$ into $\mathbf{W}$. At first glance, the time-dependent Schr\"odinger equation $i\hbar\partial_t\Psi = -(\hbar^2/2m)\partial_x^2\Psi + V\Psi$ bears no resemblance to the above equation of motion. However, upon separating into real and imaginary parts, and taking time derivatives to uncouple those parts, it transforms into precisely the form we have written above. Shen Hui-chuan gives another perspective on this analogy, which is essentially to take the square root of the WLC equation of motion,\cite{hui-chuan87} similar in spirit to how Dirac took the square root of the Klein-Gordon equation,\cite{dirac28} and to how Kane \& Lubensky took the square root of a dynamical matrix.\cite{kane14} From this perspective, the reason for introducing the substrate term is to complete a square. But again, the time-dependent analogy only holds for constant $T$, which corresponds to constant $V$, so we should be cautious in extending the claim of isomorphism to time-dependent phenomena.

Other questions one could ask include: what would be the quantum analog of a substrate term in the time-\emph{independent} buckling equation? What would be the elastic analog of an exchange term in the many-body problem? Is it possible that an intractable problem on one side can be mapped to a less difficult or more intuitive problem on the other side? This work establishes a theoretical foundation, and provides several intuition-building examples, from which further such questions can be addressed.

\appendix*   

\section{Rod buckling in 3d, and a suggested exercise}

If there is not one but two transverse dimensions into which the WLC can buckle, the deflection becomes a vector $\mathbf{u}(x)=u_1(x)\hat{y}+u_2(x)\hat{z}$. Assuming the bending modulus is isotropic, the statements of shearing force balance and inextensibility (Eqs.~(\ref{W1}) and~(\ref{W2}), respectively) are replaced with
\begin{eqnarray}
\frac{d^2\mathbf{W}}{dx^2} - \frac{T(x)}{\kappa} \mathbf{W} &=& 0,\label{SE_boldW}\\
\int dx\,\mathbf{W\cdot W} &=& 1,
\end{eqnarray}
where $\mathbf{W}(x)=W_1(x)\hat{y} + W_2(x)\hat{z} = (1/\sqrt{2\gamma L})(d\mathbf{u}/dx)$. (Compare Eq.~(\ref{SE_boldW}) to Landau \& Lifshitz's Problem 7 in Section 21 of Chapter 11, up until the point where they assume the bending modulus is \emph{anisotropic} such that only one of the transverse dimensions is relevant.\cite{landau86}) The mapping given by Eq.~(\ref{Wtopsi}) becomes
\begin{eqnarray}
W_1 \hskip 1em &\mapsto& \hskip 1em \textrm{Re}[\psi], \\
W_2 \hskip 1em &\mapsto& \hskip 1em \textrm{Im}[\psi],
\end{eqnarray}
where $\psi(x)$ is now any complex, normalized solution of the time-independent Schr\"odinger equation. Eqs.~(\ref{TtoE})--(\ref{U}) remain valid for the case at hand. Thus, a WLC that buckles in two transverse dimensions under the influence of a generalized body force $T(x)$ is \emph{isomorphic} to time-independent quantum mechanics in 1d.

As a fifth application of the mapping/isomorphism, we suggest the following multi-part exercise in which the reader can explore the buckling analog of a quantum particle on a ring. In the latter problem, the ``twisted'' boundary condition $\psi(a)=e^{i\phi}\psi(0)$, where $a$ is the ring's circumference and $\phi$ is a phase, keeps $\psi^*\psi$ continuous across the boundary. A gauge transformation can remove the twist, but at the expense of introducing a magnetic field, and this remarkable transformation is related to an underlying topology.\cite{altland06}

1. Show that the ``kinked" WLC boundary condition $\mathbf{W}(a)=R(\phi)\mathbf{W}(0)$, where $R$ is a standard rotation matrix and $\phi$ is an arbitrary angle, keeps $\mathbf{W\cdot W}$ continuous across the boundary.

2. Take $T(x)=-|T|=$ constant so that the buckling equation can be written as an eigenvalue equation
\begin{equation}
\kappa 
\begin{pmatrix}
    d/dx & 0 \\
    0 & d/dx 
\end{pmatrix}
^2
\mathbf{W} = -|T|\,\mathbf{W},
\end{equation}
with $\mathbf{W}$ a column vector. Show that substituting $\mathbf{W}(x)=R(\phi x/a){\mathbf{\tilde W}}(x)$ transforms the problem into
\begin{equation}
\kappa 
\begin{pmatrix}
    d/dx & -\phi/a \\
    \phi/a & d/dx 
\end{pmatrix}
^2
\mathbf{\tilde W} = -|T|\,\mathbf{\tilde W},
\end{equation}
where $\mathbf{\tilde W}$ has no kink. Hint: insert the identity matrix $R(\phi x/a)R(-\phi x/a)$ into a couple of strategic places.

3. To get a physical interpretation of the transformed buckling problem, map the configuration of the WLC to the trajectory of a particle moving at constant velocity $v$ in the $x$-direction, with boundary conditions that are periodic in time. Do this by putting $\tilde{W}_1\to Cy$, $\tilde{W}_2\to Cz$, and $x\to vt$, where $C$ is a scale factor to get the dimensions right. Describe the resulting physical system. If there's a magnetic field in the problem, what is its orientation?

4. Show that there are two special cases, $\phi=0$ and $\phi=\pm a\sqrt{|T|/\kappa}$, that allow the equations of motion to be easily uncoupled. Obtain an expression for the winding number $n$ in each of these cases, and solve for the eigenloads $|T_n|$. (Winding number in this context means the integer number of orbits in the $y$-$z$ plane per ring traversal).

5. The special set of \emph{unkinked} boundary conditions $\mathbf{W}(a)=R(2m\pi)\mathbf{W}(0)$, where $m=0,\pm1,\pm2,\dots$, are indistinguishable from one another in the pre-transformed problem, but the $m=0$ and $m\neq0$ versions correspond to different physical mechanisms in the transformed problem. Argue that these different mechanisms give rise to distinguishable particle trajectories, which in turn implies $m$ has observable consequences.

\begin{acknowledgments}

The author acknowledges Artur Tsobanjan, Daniel Sussman, and David Andrews for helpful comments on early drafts of this work, and two anonymous referees for queries that improved the discussion of higher dimensions and time-dependence.

\end{acknowledgments}


\begin{thebibliography}{26}

\bibitem{gautschi08} W. Gautschi, ``Leonhard Euler: his life, the man, and his works,'' 
SIAM Rev. \textbf{50} (1), 3--33 (2008).

\bibitem{taberlet17} N. Taberlet, J. Ferrand, \'E. Camus, L. Lachaud, and N. Plihon, ``How tall can gelatin towers be? An introduction to elasticity and buckling,'' 
Am. J. Phys. \textbf{85} (12), 908--914 (2017).

\bibitem{casey93} J. Casey, ``The elasticity of wood,'' 
Phys. Teach. \textbf{31}, 286--288 (1993).

\bibitem{denny19} M. Denny, ``Atlatl internal ballistics,'' 
Phys. Teach. \textbf{57}, 69--72 (2019).

\bibitem{mills60} B. D. Mills Jr., ``The fluid column,'' 
Am. J. Phys. \textbf{28}, 353--356 (1960).

\bibitem{landau86} L. D. Landau and E. M. Lifshitz, 
\textit{Theory of Elasticity}, 3rd edition (Pergamon Press, Oxford, 1986).

\bibitem{hui-chuan85} Shen Hui-chuan, ``The relation of von K\'arm\'an equation for elastic large deflection problem and Schr\"odinger equation for quantum eigenvalues problem,'' 
Appl. Math Mech.-Engl. \textbf{6} (8), 761--775 (1985).

\bibitem{hui-chuan87} Shen Hui-chuan, ``Further study of the relation of von K\'arm\'an equation for elastic large deflection problem and Schr\"odinger equation for quantum eigenvalues problem,'' 
Appl. Math Mech.-Engl. \textbf{8} (6), 561-568 (1987).

\bibitem{ohayre16} R. O'Hayre, S.-W. Cha, W. G. Colella, and F. B. Prinz, 
\textit{Fuel Cell Fundamentals}, 3rd edition (John Wiley \& Sons, Inc., 2016).

\bibitem{odijk86} T. Odijk, ``Theory of lyotropic polymer liquid crystals,'' 
Macromolecules \textbf{19} (9), 2313--2329 (1986). 

\bibitem{odijk98} T. Odijk, ``Microfibrillar buckling within fibers under compression,'' 
J. Chem. Phys. \textbf{108} (16), 6923--6928 (1998).  

\bibitem{hansen99} P. L. Hansen, D. Sven\v{s}ek, V. A. Parsegian, and R. Podgornik, ``Buckling, fluctuations, and collapse in semiflexible polyelectrolytes,'' 
Phys. Rev. E \textbf{60} (2), 1956--1966 (1999). 

\bibitem{kratky49} O. Kratky and G. Porod, ``R\"ontgenuntersuchung gel\"oster fadenmolek\"ule,'' 
Recueil des Travaux Chimiques des Pays-Bas \textbf{68} (12), 1106--1123 (1949).  

\bibitem{broedersz14} C. P. Broedersz and F. C. MacKintosh, ``Modeling semiflexible polymer networks,'' 
Rev. Mod. Phys. \textbf{86}, 995--1036 (2014).  

\bibitem{marantan18} A. Marantan and L. Mahadevan, ``Mechanics and statistics of the worm-like chain,'' 
Am. J. Phys. \textbf{86} (2), 86--94 (2018).  

\bibitem{devenica16} L. M. Devenica, C. Contee, R. Cabrejo, M. Kurek, E. F. Deveney, and A. R. Carter, ``Biophysical measurements of cells, microtubules, and DNA with an atomic force microscope,'' 
Am. J. Phys. \textbf{84} (4), 301--310 (2016).

\bibitem{kittel80} C. Kittel and H. Kroemer, 
\textit{Thermal Physics}, 2nd edition (W. H. Freeman and Company, 1980).

\bibitem{misseroni15} D. Misseroni, G. Noselli, D. Zaccaria, and D. Bigoni, ``The deformation of an elastic rod with a clamp sliding along a smooth and curved profile,'' 
Int. J. Solids Struct. \textbf{69--70}, 491--497 (2015).  

\bibitem{davies97} J. H. Davies, 
\textit{The Physics of Low-dimensional Semiconductors}, (Cambridge University Press, 1997).

\bibitem{banacloche99} J. Gea-Banacloche, ``A quantum bouncing ball,'' 
Am. J. Phys. \textbf{67} (9), 776--782 (1999).

\bibitem{stern72} F. Stern, ``Self-consistent results for $n$-type Si inversion layers,'' 
Phys. Rev. B \textbf{5} (12), 4891--4899 (1972).   

\bibitem{ashcroft76} N. W. Ashcroft and N. D. Mermin, 
\textit{Solid State Physics}, (Brooks/Cole, 1976).

\bibitem{blinder65} S. M. Blinder, ``Basic concepts of self-consistent-field theory,'' 
Am. J. Phys. \textbf{33} (6), 431--443 (1965).

\bibitem{dirac28} P. A. M. Dirac, ``The quantum theory of the electron,'' 
Proc. R. Soc. Lond. A \textbf{117}, 610--624 (1928).

\bibitem{kane14} C. L. Kane and T. C. Lubensky, ``Topological boundary modes in isostatic lattices,'' 
Nat. Phys. \textbf{10}, 39--45 (2014).

\bibitem{altland06} A. Altland and B. Simons, 
\textit{Condensed Matter Field Theory}, (Cambridge University Press, 2006).

\end{thebibliography}
\end{document}